\documentclass[a4paper,11pt]{article}

\usepackage{amsfonts,amssymb,amsmath,amsthm,dsfont,xfrac,xspace}
\usepackage{fullpage}
\usepackage{graphicx}
\usepackage{subfig}
\usepackage{cite}
\usepackage{hyperref}
\usepackage{algorithmic}
\graphicspath{{./},{fig/}}
\makeatletter
\let\NAT@parse\undefined
\makeatother
\usepackage[sort&compress, numbers]{natbib}
\usepackage{hyperref}

\newsavebox{\ieeealgbox}
\newenvironment{boxedalgorithmic}
  {\begin{lrbox}{\ieeealgbox}
   \begin{minipage}{\dimexpr\columnwidth-2\fboxsep-2\fboxrule}
   \begin{algorithmic}[1]}
  {\end{algorithmic}
   \end{minipage}
   \end{lrbox}\noindent\fbox{\usebox{\ieeealgbox}}}

\newcommand{\bsy}[1]{\boldsymbol{#1}}
\newcommand{\supp}{\text{supp}}

\DeclareMathOperator*{\argmax}{arg\,max}

\newtheorem{thm}{Theorem}

\renewcommand{\thelem}{\the \numexpr (\value{thm}+1) \relax.\arabic{lem}}

\newtheorem{lemSA}{Lemma}

\newcounter{algoCounter}

\title{On The Exact Recovery Condition of Simultaneous Orthogonal Matching Pursuit} 

\author{ Jean-Fran\c{c}ois Determe\thanks{Jean-Fran\c{c}ois Determe and Fran\c{c}ois Horlin are with the OPERA Wireless Communications Group, Universit\'e Libre de Bruxelles, 1050 Brussels, Belgium. E-mail: jdeterme@ulb.ac.be, fhorlin@ulb.ac.be. Jean-Fran\c{c}ois Determe is funded by the Belgian National Science Foundation (F.R.S.-FNRS).}  \quad J\'er\^{o}me Louveaux\footnotemark[2] \quad  Laurent Jacques\thanks{Laurent Jacques and J\'{e}r\^{o}me Louveaux are with the ICTEAM departement, Universit\'e Catholique de Louvain. E-mail: laurent.jacques@uclouvain.be, jerome.louveaux@uclouvain.be. Laurent Jacques is funded by the Belgian National Science Foundation (F.R.S.-FNRS).} \quad Fran\c{c}ois Horlin\footnotemark[1] }

\begin{document}
\maketitle

\begin{abstract}
Several exact recovery criteria (ERC) ensuring that orthogonal matching pursuit (OMP) identifies the correct support of sparse signals have been developed in the last few years. These ERC rely on the restricted isometry property (RIP), the associated restricted isometry constant (RIC) and sometimes the restricted orthogonality constant (ROC). In this paper, three of the most recent ERC for OMP are examined. The contribution is to show that these ERC remain valid for a generalization of OMP, entitled simultaneous orthogonal matching pursuit (SOMP), that is capable to process several measurement vectors simultaneously and return a common support estimate for the underlying sparse vectors. The sharpness of the bounds is also briefly discussed in light of previous works focusing on OMP.
\end{abstract}

\section{Introduction}
Recovering a high dimensional sparse signal by acquiring it through a linear measurement process returning fewer observations than its dimension is a problem often encountered in the digital signal processing literature. The field of research associated with such problems is often known to researchers as \textit{compressed sensing} or \textit{compressive sensing} (CS) \cite{donoho2006compressed}.\\

We define the support of a vector $\bsy{x} \in \mathbb{R}^n$ as $\supp (\bsy{x}) := \lbrace j \in \lbrack n \rbrack : x_j \neq 0\rbrace$ where $\lbrack n \rbrack$ denotes the set $\lbrace 1, 2, \dots, n\rbrace$ and $x_j$ denotes the $j$th entry of $\bsy{x}$. A vector is said to be $s$-sparse whenever its support exhibits a cardinality equal to or lower than $s$. 
\vspace*{-3mm}
\subsection{Signal model}

In this paper, we focus on a framework involving
\begin{enumerate}
	\item $K$ sparse signals $\bsy{x}_k \in \mathbb{R}^n$ to be recovered ($1 \leq k \leq K$),
	\item a common linear measurement process described by the matrix $\bsy{\Phi} \in \mathbb{R}^{m \times n}$,
	\item $K$ measurement vectors $\bsy{y}_k \in \mathbb{R}^m$ gathering the observations of each sparse signal when acquired through $\bsy{\Phi}$: $\bsy{y}_k = \bsy{\Phi} \bsy{x}_k$.
\end{enumerate}

To simplify the signal model, we introduce Equation (\ref{eq:sigModel}) to summarize the $K$ equations $\bsy{y}_k = \bsy{\Phi} \bsy{x}_k$ into a single one:
\begin{equation}\label{eq:sigModel}
	\bsy{Y} = \bsy{\Phi} \bsy{X}
\end{equation}
where $\bsy{Y} = \big(\bsy{y}_1, \dots, \bsy{y}_K \big) \in \mathbb{R}^{m \times K}$ and $\bsy{X} = \big(\bsy{x}_1, \dots, \bsy{x}_K \big) \in \mathbb{R}^{n \times K}$. Using this formulation, the support of $\bsy{X}$, denoted by $\supp (\bsy{X})$, is equal to the joint support $S := \cup_{k \in \lbrack K \rbrack} \supp (\bsy{x}_k)$.\\

When a model involves one measurement vector, it is referred to as a single measurement vector (SMV) model while models incorporating $K > 1$ measurement vectors are multiple measurement vector (MMV) models \cite{eldar2009robust}.\\

The columns of $\bsy{\Phi}$ are often referred to as the \textit{atoms}. This terminology being typically associated with dictionaries, it is worth emphasizing that the problem of recovering a $s$-sparse vector $\bsy{x}$ on the basis of the measurement vector $\bsy{y} = \bsy{\Phi} \bsy{x}$ is equivalent to finding $s$ columns (or atoms) of the (dictionary) matrix $\bsy{\Phi}$ that fully express $\bsy{y}$ when using the proper linear combination. The notion of atom will thus be used in the rest of this paper as it simplifies the mathematical discussions that follow.\\

We now introduce additional notions that are used afterwards. For $0 < p < \infty$ and $\bsy{x} \in \mathbb{R}^n$, we define the norms $\| \bsy{x} \|_p := (\sum_{j = 1}^n | x_j |^p)^{1/p}$ and $\| \bsy{x} \|_{\infty} := \max_{j \in \lbrack n \rbrack} |x_j|$.  In this paper, every vector should be considered as a column vector. Also, for $S \subseteq \lbrack n \rbrack$, the quantity $\bsy{x}_S$ denotes the vector formed by the entries of $\bsy{x}$ indexed by $S$. Similarly, for a matrix $\bsy{\Phi} \in \mathbb{R}^{m \times n}$, we define $\bsy{\Phi}_S$ as the matrix formed by the columns of $\bsy{\Phi}$ indexed within $S$. The Moore-Penrose pseudoinverse of any matrix $\bsy{\Phi}$ is denoted by $\bsy{\Phi}^{+}$ and its transpose is given by $\bsy{\Phi}^\mathrm{T}$.  Finally, the inner product of two vectors $\bsy{x}$ and $\bsy{y}$ is written as $\langle \bsy{x} , \bsy{y} \rangle$ and is equal to $\bsy{x}^{\mathrm{T}} \bsy{y}$.

\subsection{Simultaneous orthogonal matching pursuit}
Several algorithms exhibiting varying computational complexities have been investigated to address the problem above. For the SMV case, the greedy algorithm entitled orthogonal matching pursuit (OMP) \cite{pati1993orthogonal, davis1997adaptive}  is a classical choice because its complexity is lower than that of other algorithms such as $\ell_1$-minimization \cite{donoho2003optimally}.\\

If the $K$ sparse signals $\bsy{x}_k$ possess similar supports, \textit{i.e.}, their joint support $S := \cup_{k \in \lbrack K \rbrack} \supp (\bsy{x}_k)$ possesses a cardinality that is comparable to those of the individual supports $\supp (\bsy{x}_k)$, then it is interesting to perform a joint and common estimation of their supports \cite{gribonval2008atoms, determe2015simultaneous}. The simultaneous orthogonal matching pursuit (SOMP) algorithm \cite{tropp2006algorithms}, which is described in Algorithm~\ref{alg:SOMP}, is an extension of OMP to the MMV case and performs a joint support recovery.


\begin{figure}[!h]
	\textsc{Algorithm \refstepcounter{algoCounter}\label{alg:SOMP}\arabic{algoCounter}}:\\ 
	Simultaneous orthogonal matching pursuit (SOMP)\\
	
	\vspace{-2mm}
	\begin{boxedalgorithmic}
		\small
		\REQUIRE $\bsy{Y} \in \mathbb{R}^{m \times K}$, $\bsy{\Phi} \in \mathbb{R}^{m \times n}$, $s \geq 1$
		\STATE Initialization: $\bsy{R}^{(0)} \leftarrow \bsy{Y}$ and $S_0 \leftarrow \emptyset$
		\STATE $t \leftarrow 0$
		\WHILE{$t < s$}
		\STATE Determine the atom of $\bsy{\Phi}$ to be included in the support: \\ $j_t \leftarrow \mathrm{argmax}_{j \in \lbrack n \rbrack} ( \| (\bsy{R}^{(t)})^{\mathrm{T}} \bsy{\phi}_j \|_1 )$
		\STATE Update the support : $S_{t+1} \leftarrow S_{t} \cup \left\lbrace j_t \right\rbrace$
		\STATE Projection of each measurement vector onto $\mathrm{span}(\boldsymbol{\Phi}_{S_{t+1}})$: \\$\bsy{Y}^{(t+1)} \leftarrow \boldsymbol{\Phi}_{S_{t+1}} \boldsymbol{\Phi}_{S_{t+1}}^{+} \bsy{Y}$
		\STATE Projection of each measurement vector onto $\mathrm{span}(\boldsymbol{\Phi}_{S_{t+1}})^{\perp}$~: \\ $\bsy{R}^{(t+1)} \leftarrow \bsy{Y} - \bsy{Y}^{(t+1)}$
		\STATE $t \leftarrow t + 1$
		\ENDWHILE
		\RETURN $S_s$ \COMMENT{Support at last step}
	\end{boxedalgorithmic}
\end{figure}

As shown in Algorithm \ref{alg:SOMP}, at each iteration $t$, SOMP adds to the estimated support the index $j_t$ of the atom $\bsy{\phi}_{j_t}$ maximizing the metric $\| (\bsy{R}^{(t)})^{\mathrm{T}} \bsy{\phi}_j \|_1 = \sum_{k=1}^{K} | \langle \bsy{\phi}_{j},  \bsy{r}^{(t)}_k \rangle |$ (steps $4$ and $5$) where $\bsy{r}_k^{(t)}$ denotes the $k$th column of the residual matrix $\bsy{R}^{(t)}$. Each measurement vector $\bsy{y}_k$ is then projected onto the orthogonal complement of $\mathrm{span}(\boldsymbol{\Phi}_{S_{t+1}})$, denoted by $\mathrm{span}(\boldsymbol{\Phi}_{S_{t+1}})^{\perp}$, during steps $6$ and $7$. The algorithm terminates when the prescribed number of iterations $s$ has been reached. It is worth noticing that an atom cannot be picked twice as, once chosen, the projection onto   $\mathrm{span}(\boldsymbol{\Phi}_{S_{t+1}})^{\perp}$ ensures that $\langle \bsy{\phi}, \bsy{r}_k^{(t+1)}\rangle = 0$ if $\bsy{\phi} \in S_t$.

\subsection{Definitions}\label{subsec:definitions}

We define the concepts needed to state the results of Section \ref{sec:contrib}. First of all, the matrix $\bsy{\Phi}$ is said to satisfy the restricted isometry property (RIP) \cite{candes2006stable} of order $s$ with restricted isometry constant (RIC) $\delta_s$ (of order $s$) whenever
\begin{equation}\label{eq:defRIPRIC}
	(1 - \delta_s) \| \bsy{u} \|_2^2 \leq \| \bsy{\Phi} \bsy{u} \|_2^2 \leq (1 + \delta_s) \| \bsy{u} \|_2^2
\end{equation}
holds for all $s$-sparse vectors $\bsy{u}$. Thus, the RIP ensures that the linear operator $\bsy{\Phi}$ maintains the $\ell_2$-norm of $s$-sparse signals up to a certain extent that is quantified by means of the RIC $\delta_s$. Furthermore, if $\bsy{u}$ is supported onto $S$, the quantity $\| \bsy{\Phi} \bsy{u} \|_2^2$ is equal to $\| \bsy{\Phi}_S \bsy{u}_S \|_2^2 = \bsy{u}_S^{\mathrm{T}} (\bsy{\Phi}_S^{\mathrm{T}} \bsy{\Phi}_S) \bsy{u}_S$. The RIP therefore ensures that $ 1 - \delta_s \leq \lambda_{\mathrm{min}} ( \bsy{\Phi}_S^{\mathrm{T}} \bsy{\Phi}_S ) \leq \lambda_{\mathrm{max}} ( \bsy{\Phi}_S^{\mathrm{T}} \bsy{\Phi}_S) \leq 1 + \delta_s$ for all the supports $S$ of cardinality equal to or lower than $s$ where $\lambda_{\mathrm{min}}$ and $\lambda_{\mathrm{max}}$ denote the minimal and maximal eigenvalues, respectively. Also, it is easy to show that $\delta_s \leq \delta_{s+1}$.  \\

The $(\alpha$, $\alpha')$-restricted orthogonality constant (ROC) \cite{cai2010shifting} is defined as the smallest real number $\theta_{\alpha, \alpha'}$ for which
\begin{equation}\label{eq:defROC}
	| \langle \bsy{\Phi} \bsy{c}, \bsy{\Phi} \bsy{c}' \rangle | \leq \theta_{\alpha, \alpha'} \|\bsy{c}\|_2 \|\bsy{c}'\|_2 
\end{equation}
holds for every $\bsy{c}$, $\bsy{c}' \in \mathbb{R}^{m}$ exhibiting disjoint supports of cardinality $\alpha$ and $\alpha'$, respectively. Thus, the ROC quantifies how vectors with disjoint supports stay approximately orthogonal after projection by $\bsy{\Phi}$.\\

The ROC and the RIP are linked by the inequality \cite[Lemma 2.1]{candes2008restricted}  $\theta_{\alpha, \alpha'} \leq \delta_{\alpha + \alpha'}$ which indicates that the RIC can play a role similar to that of the ROC, albeit in a less sharp manner. Another similar inequality has been obtained in \cite[Section 2.3]{wang2012near} and is given by $\theta_{1, \alpha'} \leq \sqrt{\alpha'/(\alpha'-1)} \delta_{\alpha'}$ whenever $\alpha' \geq 2$. Another upper bound of $\theta_{1, \alpha'}$ has been recently obtained in \cite[Lemma II.3]{yang2013coherence} where the so-called $2$-coherence of the dictionnary matrix, denoted by $\nu_{\alpha'}$,  is used. They have shown \cite[Lemma II.2]{yang2013coherence} that $\nu_{\alpha'} \leq \delta_{\alpha'+1}$ so that the inequality $\theta_{1, \alpha'} \leq \nu_{\alpha'}$  is sharper than $\theta_{1, \alpha'} \leq \delta_{1 + \alpha'}$. The developments presented hereafter use the RIC-based inequality so that only the RIC intervenes in the final results. However, expressing our results using $\nu_{\alpha'}$ instead of $\delta_{1 + \alpha'}$ is straightforward. \\

Finally, it is worth defining the $\ell_{\infty}$-induced norm for matrices as $\| \bsy{\Phi} \|_{\infty \rightarrow \infty}  := \sup_{\| \bsy{\phi} \|_{\infty} = 1} \| \bsy{\Phi} \bsy{\phi} \|_{\infty}$ (where $\bsy{\Phi} \in \mathbb{R}^{m \times n}$) that can be computed as $\| \bsy{\Phi} \|_{\infty \rightarrow \infty} = \max_{i \in \lbrack m \rbrack} \sum_{j=1}^n | \phi_{i,j} |$ \cite[Lemma A.5]{foucart2013mathematical}. This quantity is interesting as it allows to write, for $A \subseteq \lbrack n \rbrack$, $\max_{j \in A} ( \| (\bsy{R}^{(t)})^{\mathrm{T}} \bsy{\phi}_j \|_1 ) = \| \bsy{\Phi}_A^{\mathrm{T}} \bsy{R}^{(t)} \|_{\infty \rightarrow \infty}$, which is reminiscent of the decision metric of SOMP. Some authors choose to write the $\ell_{\infty}$-induced norm of $\bsy{\Phi}$ as $\| \bsy{\Phi} \|_{\infty}$ but, to avoid confusions, we prefer to emphasize the distinction between the $\ell_{\infty}$-norms for vectors and matrices as both coexist in Section \ref{sec:proofs}.

\section{Contribution and related work}\label{sec:contrib}

The main contribution of this paper is to extend a recent exact recovery criterion (ERC) for OMP to its MMV counterpart, \textit{i.e.}, SOMP. An ERC  is a sufficient condition to ensure that the algorithm commits no mistake. The cornerstone of the results presented in this paper is given by Lemma~\ref{lem:RIPROCLow}.

\begin{lemSA}[A RIP and ROC-based lower bound on the maximal residual projection] \label{lem:RIPROCLow}
	Let $\bsy{X} \in \mathbb{R}^{n \times K}$ possess the support $S$. Let $\bsy{\Phi} \in \mathbb{R}^{m \times n}$ admit the RIC $\delta_{|S| }< 1$ and the $(1, |S|)$-ROC $\theta_{1, |S|} < 1$. Furthermore, $\bsy{P}^{(t)} = \bsy{\Phi}_{S_t} \bsy{\Phi}_{S_t}^+$ denotes the orthogonal projector onto $\mathrm{span}(\bsy{\Phi}_{S_t})$ where $S_t \subseteq S$, \textit{i.e.}, only correct atoms have been included to the estimated support before iteration $t$. Let $\bsy{R}^{(t)}$ be equal to $(\bsy{I} - \bsy{P}^{(t)}) \bsy{Y} =  (\bsy{I} - \bsy{P}^{(t)}) \bsy{\Phi} \bsy{X}$. Then, 
	\begin{equation}
		\dfrac{\| \bsy{\Phi}_S^{\mathrm{T}} \bsy{R}^{(t)} \|_{\infty \rightarrow \infty}}{\| \bsy{\Phi}_{\overline{S}}^{\mathrm{T}} \bsy{R}^{(t)} \|_{\infty \rightarrow \infty}} \geq \dfrac{1-\delta_{|S|}}{\theta_{1, |S|} \sqrt{|S|}}
	\end{equation}
	where $\overline{S}$ is the relative complement of $S$ with respect to $\lbrack n \rbrack$.
\end{lemSA}

Lemma~\ref{lem:RIPROCLow} establishes a lower bound on the ratio of the SOMP metric obtained for the correct atoms to that obtained for the incorrect ones. In that sense, and as it will be clarified in Theorem \ref{thm:RIPROCERC}, it straightforwardly provides an ERC guaranteeing that SOMP commits no error when picking atoms. We now propose a corollary of Lemma~\ref{lem:RIPROCLow} that only relies on the RIC.
\begin{lemSA}[RIP lower bounds on the maximal residual projection] \label{lem:RIPLowBounds}
	Let $\bsy{X} \in \mathbb{R}^{n \times K}$ possess the support $S$. Let $\bsy{\Phi} \in \mathbb{R}^{m \times n}$ admit the RIC $\delta_{|S| }< 1$. Furthermore, $\bsy{P}^{(t)} = \bsy{\Phi}_{S_t} \bsy{\Phi}_{S_t}^+$ denotes the orthogonal projector onto $\mathrm{span}(\bsy{\Phi}_{S_t})$ where $S_t \subseteq S$, \textit{i.e.}, only correct atoms have been included to the estimated support before iteration $t$. Let $\bsy{R}^{(t)}$ be equal to $(\bsy{I} - \bsy{P}^{(t)}) \bsy{Y} =  (\bsy{I} - \bsy{P}^{(t)}) \bsy{\Phi} \bsy{X}$. Then, both inequalities below hold
	\begin{equation}\label{eq:RIPLowBound1}
		\dfrac{\| \bsy{\Phi}_S^{\mathrm{T}} \bsy{R}^{(t)} \|_{\infty \rightarrow \infty}}{\| \bsy{\Phi}_{\overline{S}}^{\mathrm{T}} \bsy{R}^{(t)} \|_{\infty \rightarrow \infty}} \geq \dfrac{1-\delta_{|S|+1}}{\delta_{|S| +1} \sqrt{|S|}}
	\end{equation}
	\begin{equation}\label{eq:RIPLowBound2}
		\dfrac{\| \bsy{\Phi}_S^{\mathrm{T}} \bsy{R}^{(t)} \|_{\infty \rightarrow \infty}}{\| \bsy{\Phi}_{\overline{S}}^{\mathrm{T}} \bsy{R}^{(t)} \|_{\infty \rightarrow \infty}} \geq \dfrac{(1 - \delta_{|S|}) \sqrt{|S|-1}}{\delta_{|S|} |S|}
	\end{equation}
	where $\overline{S}$ is the relative complement of $S$ with respect to $\lbrack n \rbrack$.
\end{lemSA}

Compared to former works that directly derived the ERC \cite{wang2012improved, dan2014robustness, mo2012remark, wang2012recovery, wang2012near}, we believe that Lemma~\ref{lem:RIPROCLow} and Lemma~\ref{lem:RIPLowBounds} are interesting as they quantify the robustness of the decisions made at each iteration of SOMP in the noiseless case. Such quantities can then be used to produce theoretical analyses of greedy algorithms in a noisy setting (see \cite{cai2011orthogonal, dan2014robustness, determe2015simultaneous}). Other similar works analysing OMP in a noisy environment include \cite{shen2015sparse} and \cite{yang2013coherence}. The analysis presented in \cite{dan2014robustness} actually uses a result fundamentally identical to Lemma \ref{lem:RIPROCLow} for $K = 1$ to conduct a theoretical analysis of OMP inspired from  \cite{cai2011orthogonal}. We now state the three ERC deriving from Lemma~\ref{lem:RIPROCLow} and Lemma~\ref{lem:RIPLowBounds}.

\begin{thm}[Several RIP and ROC-based ERC for SOMP] \label{thm:RIPROCERC}
	Let $\bsy{X} \in \mathbb{R}^{n \times K}$ possess the support $S$. Let $\bsy{\Phi} \in \mathbb{R}^{m \times n}$ admit the RIC $\delta_{|S| }< 1$ and the $(1, |S|)$-ROC $\theta_{1, |S|} < 1$.  Then, SOMP commits no error and identifies the full support of $\bsy{X}$ at the end of iteration $|S| - 1$ whenever at least one of the three conditions below hold:
	\begin{equation}\tag{ERC1}\label{eq:ERC1}
		\dfrac{1-\delta_{|S|}}{\theta_{1, |S|} \sqrt{|S|}} > 1
	\end{equation}
	\begin{equation}\tag{ERC2}\label{eq:ERC2}
		\delta_{|S|+1} < \dfrac{1}{\sqrt{|S|} + 1}
	\end{equation}
	\begin{equation}\tag{ERC3}\label{eq:ERC3}
		(\mathrm{for } |S| \geq 2) \;\; \delta_{|S|} < \dfrac{\sqrt{|S| - 1}}{\sqrt{|S| - 1} + |S|}.
	\end{equation}
\end{thm}

As demonstrated in Section~\ref{sec:proofMainThm}, Theorem~\ref{thm:RIPROCERC} is a straightforward consequence of Lemma~\ref{lem:RIPROCLow} and Lemma~\ref{lem:RIPLowBounds}. The authors of \cite{wang2012improved} and \cite{dan2014robustness} independently obtained (\ref{eq:ERC1}) for OMP. To the best of the authors' knowledge, the second ERC was first obtained simultaneously in \cite{mo2012remark} and \cite{wang2012recovery} while (\ref{eq:ERC3}) was initially published in \cite{wang2012near}, both ERC being derived for OMP.\\

Regarding older works, it is also worth pointing out that the ERC $\delta_{|S|+1} < 1/((1 + \sqrt{2}) \sqrt{|S|})$, first obtained in \cite[Theorem 5.2]{liu2012orthogonal} for OMP, has been shown to remain valid for SOMP in \cite[Corollary 1]{ding2012robustness}.  Thereby, the authors of \cite{ding2012robustness} also proved that the older ERC $\delta_{|S|+1} < 1/(3 \sqrt{|S|})$, initially derived in \cite[Theorem 3.1]{davenport2010analysis} for OMP, remains correct for SOMP as $\delta_{|S|+1} < 1/(3 \sqrt{|S|})$ is implied by $\delta_{|S|+1} < 1/((1 + \sqrt{2}) \sqrt{|S|})$. Very recently, (\ref{eq:ERC2}) was extended to SOMP in \cite[Remark 1]{xu2015perturbation}. However, the extension to SOMP of both (\ref{eq:ERC1}) and (\ref{eq:ERC3}) is a novel result. In \cite{mo2015sharp}, the author has derived the ERC $\delta_{|S|+1} < 1/\sqrt{|S|+1}$, which is sharper than (\ref{eq:ERC2}). Combining the ideas developed in \cite{mo2015sharp} and our paper could possibly extend this ERC to SOMP.  \\

Finally, we would like to point out that, if any of the considered ERC holds, running $K$ independent executions of OMP instead of a single instance of SOMP would enable one to retrieve the individual supports $\mathrm{supp}(\bsy{x}_k)$ ($1 \leq k \leq K$) and, by extension, the joint support $S$. While it may seem to undermine the interest of this work, the following observations convince otherwise:
\begin{enumerate}
	\item If one of the considered ERC guarantees that each one of the $K$ instances of OMP returns the correct support of each sparse vector $\bsy{x}_k$, then SOMP is also guaranteed to return the correct joint support so that there is no penalty switching from OMP to SOMP, except maybe that SOMP returns a joint support instead of possibly smaller (yet correct) supports for each $\bsy{x}_k$.
	\item Lemma \ref{lem:RIPROCLow} and Lemma \ref{lem:RIPLowBounds} should be thought of as the central results of this paper as they quantify the robustness of the support recovery in the noiseless case, the resulting ERC being merely direct consequences of the aforementioned lemmas. As mentioned previously, these lemmas can be used to produce theoretical analyses of SOMP for noisy scenarios while it is not the case for the ERC.
\end{enumerate}

\section{Sharpness of the bounds}

In \cite{dan2013sharp}, it is shown that (\ref{eq:ERC1}) is sharp for OMP in the sense that it is possible to construct a measurement matrix $\bsy{\Phi}_{\mathrm{bad}}$ satisfying $(1-\delta_{|S|})/(\theta_{1, |S|} \sqrt{|S|}) = 1$ for which there exists a $|S|$-sparse signal $\bsy{x}_{\mathrm{bad}}$ that OMP fails to recover. The sharpness property is immediately extended to SOMP by noticing that if OMP fails to recover $\bsy{x}_{\mathrm{bad}}$ on the basis of the measurement vector $\bsy{y}_{\mathrm{bad}} = \bsy{\Phi}_{\mathrm{bad}} \bsy{x}_{\mathrm{bad}}$, then SOMP also fails with $\bsy{Y}_{\mathrm{bad}} = \bsy{\Phi}_{\mathrm{bad}} \bsy{X}_{\mathrm{bad}}$ where $\bsy{X}_{\mathrm{bad}} = \big( \bsy{x}_{\mathrm{bad}}, \dots, \bsy{x}_{\mathrm{bad}} \big)$ as both algorithms make the same decisions in this case.\\

Regarding (\ref{eq:ERC2}) and (\ref{eq:ERC3}), it has been shown  in \cite{mo2015sharp} that there exists a signal $\bsy{x}_{\mathrm{bad}}$ of support $S$ and a matrix $\bsy{\Phi}_{\mathrm{bad}}$ satisfying  $\delta_{|S|+1} = 1 / \sqrt{|S|+1}$ for which OMP fails to recover the support of $\bsy{x}_{\mathrm{bad}}$ on the basis of $\bsy{y}_{\mathrm{bad}} = \bsy{\Phi}_{\mathrm{bad}} \bsy{x}_{\mathrm{bad}}$. Note that earlier works (see \cite{mo2012remark} and \cite{wang2012recovery}) proved that the statement above holds with $\bsy{\Phi}_{\mathrm{bad}}$ satisfying $\delta_{|S|+1} = 1 / \sqrt{|S|}$. Using an approach identical to that of the previous paragraph, one shows that this statement remains true for SOMP with $\bsy{Y}_{\mathrm{bad}} = \bsy{\Phi}_{\mathrm{bad}} \bsy{X}_{\mathrm{bad}}$ and $\bsy{X}_{\mathrm{bad}} = \big( \bsy{x}_{\mathrm{bad}}, \dots, \bsy{x}_{\mathrm{bad}} \big)$. It shows that (\ref{eq:ERC2}) is near-optimal as, for $|S| \rightarrow \infty$, it boils down to the condition $\delta_{|S|+1} < 1 / \sqrt{|S|+1}$. It can be shown that (\ref{eq:ERC3}) is also near-optimal but the discussion is more involved as $\delta_{|S|}$ intervenes instead of $\delta_{|S|+1}$. In \cite[Section 3]{wang2012near}, it is shown that $\delta_{|S|+1} < 1/(|S| + 3 - \sqrt{2})$ implies (\ref{eq:ERC3}), therefore indicating that (\ref{eq:ERC3}) is also at least near-optimal.

\section{Proofs}\label{sec:proofs}
\subsection{Proof of Lemma \ref{lem:RIPROCLow}}

The proof presented in this section is analog to what has been proposed in \cite{wang2012improved, wang2012near, dan2014robustness}, the only difference being the additional quantities needed to deal with the MMV model. \\

The proof is decomposed in three steps:
\begin{enumerate}
	\item Derive an upper bound on $\| \bsy{\Phi}_{\overline{S}}^{\mathrm{T}} \bsy{R}^{(t)} \|_{\infty \rightarrow \infty}$ expressed as $\theta_{1, |S|} \| \bsy{z}^{(t)} \|_2$ where $\bsy{z}^{(t)}$ is to be specified in the detailed development.
	\item Derive a lower bound on $\| \bsy{\Phi}_{S}^{\mathrm{T}} \bsy{R}^{(t)} \|_{\infty \rightarrow \infty}$ expressed as $(1/\sqrt{|S|}) (1 - \delta_{|S|}) \| \bsy{z}^{(t)} \|_2$ where $\bsy{z}^{(t)}$ is identical for steps $1$) and $2$).
	\item Compute the ratio of the lower bound to the upper bound and observe that the desired result is obtained due to the cancellation of the quantity $\| \bsy{z}^{(t)} \|_2$.
\end{enumerate}

Let us first tackle the quantity $ \| \bsy{\Phi}_{\overline{S}}^{\mathrm{T}} \bsy{R}^{(t)} \|_{\infty \rightarrow \infty} = \max_{j \in \overline{S}} ( \sum_{k=1}^{K} | \langle \bsy{r}_{k}^{(t)}, \bsy{\phi}_j \rangle | )$ and define $j^*(t) := \argmax_{j \in \overline{S}} ( \sum_{k=1}^{K} | \langle \bsy{r}_{k}^{(t)}, \bsy{\phi}_j \rangle | )$. Then, if $c_k^{(t)} := \mathrm{sign} ( \langle \bsy{r}_k^{(t)}, \bsy{\phi}_{j^*(t)} \rangle )$, we have 
\begin{align*}
	\| \bsy{\Phi}_{\overline{S}}^{\mathrm{T}} \bsy{R}^{(t)} \|_{\infty \rightarrow \infty} & = \max_{j \in \overline{S}} \left( \sum_{k=1}^{K} | \langle \bsy{r}_{k}^{(t)}, \bsy{\phi}_j \rangle | \right) \\
	& =  \left| \sum_{k=1}^{K} c_k^{(t)} \langle \bsy{r}_{k}^{(t)}, \bsy{\phi}_{j^*(t)} \rangle \right|  \\
	& =    \left| \left\langle \sum_{k=1}^{K} c_k^{(t)} \bsy{r}_{k}^{(t)}, \bsy{\phi}_{j^*(t)} \right\rangle \right|. \\
\end{align*}
Since $S_t \subseteq S$, $\bsy{r}_k^{(t)} = (\bsy{I} - \bsy{P}^{(t)}) \bsy{r}_k$ belongs to $\mathrm{span} (\bsy{\Phi}_S)$ and can thus be expressed as a linear combination of the atoms whose indexes belong to $S$ by means of $\bsy{r}_k^{(t)} = \bsy{\Phi}_S \bsy{a}_k^{(t)}$ where  $\bsy{a}_k^{(t)} \in \mathbb{R}^{|S|}$ contains the coefficients of the linear combination of interest. It is also worth defining the extension $\bsy{\tilde{a}}_k^{(t)}$ of $\bsy{a}_k^{(t)}$ to $\mathbb{R}^{n}$ by ensuring that $\supp (\bsy{\tilde{a}}_k^{(t)}) \subseteq S$ and $( \bsy{\tilde{a}}_k^{(t)} )_S = \bsy{a}_k^{(t)}$. Another relation of interest is $\bsy{\phi}_{j^*(t)} = \bsy{\Phi} \bsy{e}_{j^*(t)}$ where $\bsy{e}_{j^*(t)}$ denotes the $j^*(t)$th vector of the canonical basis of $\mathbb{R}^n$. Hence, using consecutively the equations of this paragraph and the definition of the ROC yields
\begin{align*}
	\| \bsy{\Phi}_{\overline{S}}^{\mathrm{T}} \bsy{R}^{(t)} \|_{\infty \rightarrow \infty} & =  \left| \left\langle \bsy{\Phi} \sum_{k=1}^{K} c_k^{(t)} \bsy{\tilde{a}}_{k}^{(t)}, \bsy{\Phi} \bsy{e}_{j^*(t)} \right\rangle \right| \\
	& \leq \theta_{1, |S|} \left\| \sum_{k=1}^{K} c_k^{(t)} \bsy{\tilde{a}}_{k}^{(t)} \right\|_2
\end{align*}
where $\| \bsy{e}_{j^*(t)} \|_2$ is equal to $1$. It is worth explicitly pointing out that the ROC definition is applicable in that case because the supports of $\bsy{e}_{j^*(t)}$ and $\sum_{k=1}^{K} c_k^{(t)} \bsy{\tilde{a}}_{k}^{(t)}$  are disjoint as $j^*(t) \in \overline{S}$ and $\supp (\bsy{\tilde{a}}_{k}^{(t)}) \subseteq S$ for $1 \leq k \leq K$.\\

The first step of the proof is now completed and the last problem to be dealt with is deriving a lower bound for $\| \bsy{\Phi}_{S}^{\mathrm{T}} \bsy{R}^{(t)} \|_{\infty \rightarrow \infty}$. For any $d_k^{(t)} \in \lbrace -1; 1\rbrace$, we have $|\langle \bsy{r}_k^{(t)}, \bsy{\phi}_j \rangle| = 
|d_k^{(t)} \langle \bsy{r}_k^{(t)}, \bsy{\phi}_j \rangle| = | \langle d_k^{(t)} \bsy{r}_k^{(t)}, \bsy{\phi}_j \rangle|$. In particular, it remains true for the choice $d_k^{(t)} = c_k^{(t)}$. Thus, by using the equation above and the triangle inequality, one obtains
\begin{align*}
	\| \bsy{\Phi}_{S}^{\mathrm{T}} \bsy{R}^{(t)} \|_{\infty \rightarrow \infty} & = \max_{j \in S} \left( \sum_{k=1}^{K} | \langle \bsy{r}_{k}^{(t)}, \bsy{\phi}_j \rangle | \right) \\
	& = \max_{j \in S} \left( \sum_{k=1}^{K} | \langle c_k^{(t)} \bsy{r}_{k}^{(t)}, \bsy{\phi}_j \rangle | \right) \\
	& \geq \max_{j \in S} \left| \left\langle \sum_{k=1}^{K} c_k^{(t)} \bsy{r}_{k}^{(t)}, \bsy{\phi}_j \right\rangle \right| \\
	& = \left\| \bsy{\Phi}_S^{\mathrm{T}} \left( \sum_{k=1}^{K} c_k^{(t)} \bsy{r}_{k}^{(t)} \right) \right\|_{\infty} \\
	& \geq \dfrac{1}{\sqrt{|S|}} \left\| \bsy{\Phi}_S^{\mathrm{T}} \left( \sum_{k=1}^{K} c_k^{(t)} \bsy{r}_{k}^{(t)} \right) \right\|_{2}
\end{align*}
where $\bsy{\Phi}_S^{\mathrm{T}} ( \sum_{k=1}^{K} c_k^{(t)} \bsy{r}_{k}^{(t)} ) \in \mathbb{R}^{|S|}$. Also, we have previously obtained $\bsy{r}_k^{(t)} = \bsy{\Phi}_S \bsy{a}_k^{(t)}$. The lower bound on $\| \bsy{\Phi}_{S}^{\mathrm{T}} \bsy{R}^{(t)} \|_{\infty \rightarrow \infty}$ is thus finally obtained by successively using the two previous relations and the inequality $1 - \delta_{|S|} \leq \lambda_{\mathrm{min}} ( \bsy{\Phi}_S^{\mathrm{T}} \bsy{\Phi}_S )$ resulting from the RIP (see Section \ref{subsec:definitions}) in the following manner:
\begin{align*}
	\| \bsy{\Phi}_{S}^{\mathrm{T}} \bsy{R}^{(t)} \|_{\infty \rightarrow \infty} & \geq \dfrac{1}{\sqrt{|S|}} \left\| \bsy{\Phi}_S^{\mathrm{T}} \bsy{\Phi}_S \left( \sum_{k=1}^{K} c_k^{(t)} \bsy{a}_{k}^{(t)}  \right)  \right\|_2 \\
	& \geq \dfrac{1 - \delta_{|S|}}{\sqrt{|S|}} \left\| \sum_{k=1}^{K} c_k^{(t)} \bsy{a}_{k}^{(t)}   \right\|_2
\end{align*}
where $\| \sum_{k=1}^{K} c_k^{(t)} \bsy{a}_{k}^{(t)} \|_2 = \| \sum_{k=1}^{K} c_k^{(t)} \bsy{\tilde{a}}_{k}^{(t)} \|_2$. The final result is now established by expressing the ratio of the lower bound on $\| \bsy{\Phi}_{S}^{\mathrm{T}} \bsy{R}^{(t)} \|_{\infty \rightarrow \infty}$ to the upper bound on $\| \bsy{\Phi}_{\overline{S}}^{\mathrm{T}} \bsy{R}^{(t)} \|_{\infty \rightarrow \infty}$. \qed

\subsection{Proof of Lemma \ref{lem:RIPLowBounds}}

The proof consists in finding lower bounds on the ratio $(1-\delta_{|S|})/(\theta_{1, |S|} \sqrt{|S|})$ intervening in Lemma~\ref{lem:RIPROCLow}. For the first bound, it is sufficient to use the inequalities $\delta_{|S|} \leq \delta_{|S|+1}$ and $\theta_{1, |S|} \leq \delta_{|S|+1}$ \cite[Lemma 2.1]{candes2008restricted}, for the numerator and the denominator respectively. The second bound is obtained by using the inequality $\theta_{1, |S|} \leq \sqrt{|S|/(|S|-1)} \delta_{|S|}$ on the denominator for $|S| \geq 2$ \cite[Section 2.3]{wang2012near}. \qed

\subsection{Proof of Theorem \ref{thm:RIPROCERC}}\label{sec:proofMainThm}

Let us first address the proof of (\ref{eq:ERC1}). At iteration $0$, we have $\bsy{R}^{(t)} = \bsy{Y}$ and Lemma~\ref{lem:RIPROCLow} shows that a sufficient condition for SOMP to pick a correct atom is $(1-\delta_{|S|})/(\theta_{1, |S|} \sqrt{|S|}) > 1$ as it means that the highest metric is necessarily obtained for one of the correct atoms. Thus, at iteration $1$, the condition $S_1 \subseteq S$ is verified and Lemma~\ref{lem:RIPROCLow} shows, once again, that a correct decision will be made. By repeatedly applying the same train of thought, one proves the theorem by induction. The remaining ERC are obtained in an identical manner by using the two bounds provided by Lemma~\ref{lem:RIPLowBounds} instead of that of Lemma~\ref{lem:RIPROCLow}. \qed

\section*{Acknowledgments}

The authors would like to thank the Belgian ``Fonds de la recherche scientifique'' for having funded this research.

\newpage
\nocite{*}
\bibliographystyle{abbrv}
\bibliography{mybib}

\end{document}